\documentclass[
    aps,
    prl,
    twocolumn,
    10pt,
    a4paper,
    superscriptaddress,
    notitlepage,
    longbibliography,
    floatfix
]{revtex4-2}

\DeclareMathAlphabet      {\mathbfit}{OML}{cmm}{b}{it}

\usepackage{titlesec}

\titlespacing*{\section}{0pt}{2.0ex}{2.0ex}     
\titlespacing*{\subsection}{0pt}{1.5ex}{1.5ex}

\usepackage{array}
\usepackage{rotating}
\usepackage{braket}

\usepackage[version=4]{mhchem}
\usepackage{placeins}
\usepackage{tabularx}
\usepackage{xcolor}
\definecolor{navyblue}{RGB}{0,0,128}
\definecolor{lightgrey}{gray}{0.65}
\usepackage{amsmath}
\usepackage{multirow}
\usepackage{colortbl,xcolor}
\usepackage{longtable}
\usepackage{graphicx}
\usepackage{bm}
\usepackage{siunitx}
\usepackage{afterpage}
\usepackage{tikz}
\usepackage{hyperref} 
\usepackage{booktabs}
\usepackage{multirow}
\usepackage[normalem]{ulem}
\usepackage{makecell} 
\usepackage{amssymb} 

\usepackage{pifont}

\usepackage{newtxtext,newtxmath}

\renewcommand{\vec}[1]{\boldsymbol{#1}}

\usepackage[caption=false]{subfig}
\usepackage[normalem]{ulem}
\usepackage{xcolor}


\usepackage{orcidlink}
\usepackage{hyperref}
\hypersetup{colorlinks=true, citecolor=blue, urlcolor=blue, linkcolor=blue}

\begin{document}
\title{Anti-spin Laue groups: 
classification of anti-altermagnets and their
representative minimal models}

\author{Colin Lange \orcidlink{0009-0001-0661-7998}}
\affiliation{Institute of Physics, Johannes Gutenberg University Mainz, 55099 Mainz, Germany}
\author{Rodrigo Jaeschke-Ubiergo \orcidlink{0000-0002-4821-8303}}
\affiliation{Institute of Physics, Johannes Gutenberg University Mainz, 55099 Mainz, Germany}
\author{Alexander Mook \orcidlink{0000-0002-8599-9209}}
\affiliation{University of M\"{u}nster, Institute of Solid State Theory, 48149 Münster, Germany}
\author{Jairo Sinova \orcidlink{0000-0002-9490-2333}}
\affiliation{Institute of Physics, Johannes Gutenberg University Mainz, 55099 Mainz, Germany}
\affiliation{Department of Physics, Texas A\&M University, College Station, Texas 77843-4242, USA}

\begin{abstract}
Anti-altermagnets exhibit odd-parity nonrelativistic spin splitting, yet unlike even-parity altermagnets, their momentum-space symmetries lack a reduced classification analogous to spin Laue groups. Here, we introduce \textit{anti-spin Laue groups}, organized into three distinct classes, and identify 21 groups describing the odd-parity partial wave character of this unconventional class. Together with the 10 spin Laue groups of altermagnets, they complete the classification of nonrelativistic unconventional magnets with collinear momentum-space spin polarization. Anti-spin Laue groups also provide a many-to-one reduction of spin space (point) groups by retaining only their action on the collinear momentum-space spin polarization, thereby directly encoding the symmetry-enforced nodal spin-splitting character. Based on this we develop a systematic model-construction algorithm yielding minimal, material-oriented four-band models. This framework places odd- and even-parity unconventional magnets on equal footing within a unified momentum-space symmetry description.
\end{abstract}
\maketitle

\noindent{\it Introduction} -- Altermagnetism is a novel collinear magnetic phase that combines nonrelativistic spin-split electronic band structure with vanishing net magnetization~\cite{Smejkal2021a, Smejkal2022a,Mazin2022a,Jungwirth2025a}. Its unconventional spin splitting is even under parity and exhibits characteristic higher-partial-wave anisotropy, denoted as $d$-, $g$-, and $i$-wave~\cite{Smejkal2021a}. The experimental verification of bulk altermagnets, including $\mathrm{MnTe}$ and $\mathrm{CrSb}$~\cite{Krempasky2024,Lee2024,Osumi2024,Hajlaoui2024,Chilcote2024,Amin2024,Reimers2024,Yang2024,Ding2024,Zeng2024,Li2024,Liu2024b}, together with the development of symmetry-based minimal models~\cite{Roig2024,Fernandes2023,Ezawa2025a}, has stimulated predictions of a broad range of transport and magneto-optical phenomena \cite{Jungwirth2026,Smejkal2020,Mazin2021,Gonzalez-Hernandez2021,Samanta2020, Smejkal2022GMR,Bai2024,Shao2021,Zhu2023a,Jaeschke-Ubiergo2023,Karetta2025,Jaeschke-Ubiergo2025,Trama2024,Golub2025,Jaeschke-Ubiergo2026}, with abundant experimental evidence \cite{Feng2022,Betancourt2021,Reichlova2024,Leiviska2024,Han2024,Galindez-Ruales2025a,Galindez-Ruales2025,Badura2025,Bose2022,Bai2022,Karube2022,Liu2023,Liao2024,Hariki2023,Fedchenko2024}. 

\begin{table}[t]
  \centering
  \caption{\textbf{Three distinct types of (anti-)spin Laue groups ((A)SLGs) under even- and odd-parity constraints.}
  $\mathbf{R}_{\mathrm{s}}^{\mathrm{I}}$, $\mathbf{R}_{\mathrm{s}}^{\mathrm{II}}$, and
  $\mathbf{R}_{\mathrm{s}}^{\mathrm{III}}$ denote the different classes of Laue groups.
  The table shows the types of magnets described by these Laue groups.
  $S_\parallel$ denotes an in-plane spin-polarization component of a coplanar magnet,
  and EPM denotes an even-parity magnet.
  FM, AFM, and AM stand for ferro-, antiferro-, and altermagnet, respectively.
  AAM denotes anti-altermagnet and OPM odd-parity magnet.}
  \label{tab:spin_laue_group_types}

  \setlength{\tabcolsep}{3pt}
  \renewcommand{\arraystretch}{1.18}

  \begin{tabular}{lccc}
    \toprule
      & $\mathbf{R}_{\mathrm{s}}^{\mathrm{I}}$
      & $\mathbf{R}_{\mathrm{s}}^{\mathrm{II}}$
      & $\mathbf{R}_{\mathrm{s}}^{\mathrm{III}}$ \\
    \midrule
    \makecell[l]{Even-parity \\ SLG, collinear}
      & \makecell[c]{ 
      FM}
      & \makecell[c]{
      AFM}
      & \makecell[c]{ 
      AM} \\
    \midrule
    \makecell[l]{Even-parity\\ SLG, coplanar}
      & \makecell[c]{$S_\parallel\neq 0$\\nodeless}
      & \makecell[c]{$S_\parallel= 0$\\ 
      AFM}
      & \makecell[c]{$S_\parallel\neq 0$\\nodal EPM} \\
    \midrule
    \makecell[l]{Odd-parity\\anti-SLG, coplanar}
      & ---
      & \makecell[c]{$S_\perp =0$\\
      AFM}
      & \makecell[c]{AAM/OPM} \\
    \bottomrule
  \end{tabular}
\end{table}

\begin{table}[t]
  \centering
  \caption{\textbf{Anti-spin Laue groups describing odd-parity magnets} The spin splitting column shows a schematic of the spin-splitting texture in momentum space together with the partial wave character. P-$f$ and B-$f$ represent the planar and bulk like $f$-wave groups, in analogy to the notation used in Ref. \cite{Smejkal2021a}. The second column shows the anti-spin Laue groups that are constructed with the halving subgroups of the third column. The number shown in the fourth column shows how many physically non equivalent spin point groups correspond to the respective anti-spin Laue group. The final column gives some representatives of the 83 material candidates obtained in our MAGNDATA screening, when available.}
  \renewcommand{\arraystretch}{1.22}
  \setlength{\tabcolsep}{2pt}

  \begin{tabular}{c|c|c|c|c}
    \makecell[c]{$S_\perp(\boldsymbol{k})$}
      & \makecell[c]{Anti-spin\\Laue group}
      & \makecell[c]{Halving\\subgroup}
      & \makecell[c]{$N_{252}$}
      & \makecell[c]{Material\\Candidate}\\
    \hline

    \multirow{10}{*}{\makecell[c]{\includegraphics[width=0.82cm]{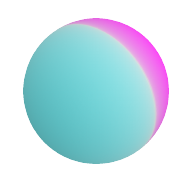}\\$p$-wave}}
      & ${}^{2}\bar{1}$ & $1$ & 2 & NiI$_2$ \\
      & ${}^{1}2/{}^{2}m$ & $2$ & 5 & NiPS$_3$ \\
      & ${}^{2}2/{}^{1}m$ & $m$ & 5 & CeNiAsO \\
      & ${}^{2}m{}^{1}m{}^{1}m$ & $mm2$ & 10 & Ca$_2$Cr$_2$O$_5$ \\
      & ${}^{1}4/{}^{2}m$ & $4$ & 6 & ---\\
      & ${}^{1}4/{}^{2}m{}^{1}m{}^{1}m$ & $4mm$ & 11 & ---\\
      & ${}^{2}\bar{3}$ & $3$ & 3 & KFe(PO$_3$F)$_2$ \\
      & ${}^{2}\bar{3}{}^{1}m$ & $32$ & 6 & GeCo$_2$O$_4$ \\
      & ${}^{1}6/{}^{2}m$ & $6$ & 7 & --- \\
      & ${}^{1}6/{}^{2}m{}^{1}m{}^{1}m$ & $622$ & 12 & --- \\
    \hline

    \multirow{5}{*}{\makecell[c]{\includegraphics[width=0.82cm]{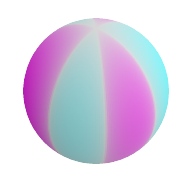}\\$\mathrm{P}$-$f$}}
      & & & & \\
      & ${}^{2}\bar{3}{}^{2}m$ & $3m$ & 6 & InMnO$_3$ \\
      & ${}^{2}6/{}^{1}m$ & $\bar{6}$ & 7 & TmPdIn \\
      & ${}^{2}6/{}^{1}m{}^{1}m{}^{2}m$ & $\bar{6}m2$ & 16 & ErAuIn\\
      & & & & \\
    \cline{1-5}

    \multirow{7}{*}{\makecell[c]{\includegraphics[width=0.82cm]{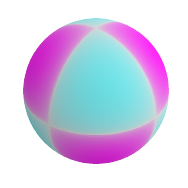}\\$\mathrm{B}$-$f$}}
      & & & & \\
      & ${}^{2}m{}^{2}m{}^{2}m$ & $222$ & 6 & LuMnO$_3$ \\
      & ${}^{2}4/{}^{2}m{}^{2}m{}^{1}m$ & $\bar{4}2m$ & 14 & Sr$_2$FeO$_3$Cl\\
      & ${}^{2}m{}^{2}\bar{3}$ & $23$ & 3 & VCl$_2$ \\
      & ${}^{2}m{}^{2}\bar{3}{}^{1}m$ & $\bar{4}3m$ & 6 & ---\\
      & & & & \\
    \hline

    \makecell[c]{\includegraphics[width=0.82cm]{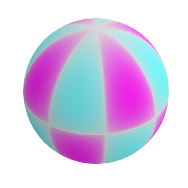}\\$h$-wave}
      & ${}^{1}4/{}^{2}m{}^{2}m{}^{2}m$ & $422$ & 11 & LaMnAu$_5$\\
    \hline

    \makecell[c]{\includegraphics[width=0.82cm]{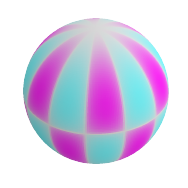}\\$j$-wave}
      & ${}^{1}6/{}^{2}m{}^{2}m{}^{2}m$ & $6mm$ & 12 & ---\\
    \hline

    \makecell[c]{\includegraphics[width=0.82cm]{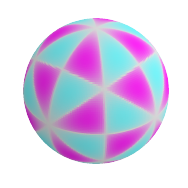}\\$l$-wave}
      & ${}^{2}m{}^{2}\bar{3}{}^{2}m$ & $432$ & 6 & --- \\
  \end{tabular}

  \label{tab:onecolumn_spin_splitting}
\end{table}

The underlying symmetry framework is provided by spin groups, which decouple spin and real-space operations in the nonrelativistic limit~\cite{Litvin1977,Litvin1974,Brinkman1966,Watanabe2024,Schiff2025,Xiao2024}. For collinear magnets, the momentum-space symmetry is naturally captured by \emph{spin Laue groups}, which augment the spin group by effective real-space inversion and provide a complete classification of even-parity spin splitting~\cite{Smejkal2021a,Schiff2025a}. 

More recently, the concept of unconventional magnetism has been extended to compensated noncollinear magnets with odd-parity nonrelativistic spin splitting~\cite{Hellenes2023,Chakraborty2025a,Mitscherling2026}. In coplanar magnets with $\mathcal{T}\vec{t}$ symmetry, where $\mathcal{T}$ and $\vec{t}$ denote time reversal symmetry and a translation, respectively, the in-plane spin expectation values of Bloch electrons vanish identically in the non-relativistic limit, leaving a collinear spin polarization which is perpendicular to coplanar magnetic moments. This remaining component is odd under momentum inversion,
$
S_\perp(-\vec{k})=-S_\perp(\vec{k}),
$
giving rise to characteristic odd parity ($p$-, $f$-, $h$-, $\dots$)-wave spin-splitting patterns. We refer to these coplanar magnets with $\mathcal{T}\vec{t}$ symmetry as anti-altermagnets, being the
odd-parity counterparts of altermagnets~\cite{Hellenes2023,Hellenes2023a,Jungwirth2025}.

Despite the rapid development of the field~\cite{Brekke2024,Song2025b,Chakraborty2025a,Yamada2025}, anti-altermagnets still lack a reduced momentum-space symmetry description analogous to spin Laue groups. Existing classifications determine the allowed spin splitting for individual spin point or spin space groups~\cite{Luo2025,Elcoro2026}, faithfully characterizing the underlying magnetic order but not directly exposing the common momentum-space symmetry shared by many such groups.

In this work, we introduce \emph{anti-spin Laue groups}, as the minimal symmetry objects governing the odd-parity spin-splitting in coplanar magnets with collinear spin polarization in momentum space. In particular, they provide a symmetry classification of anti-altermagnets. By retaining only whether a spin symmetry preserves or reverses the spin component perpendicular to the real space coplanar order, anti-spin Laue groups provide the natural odd-parity counterpart of spin Laue groups and place even- and odd-parity compensated magnets on equal footing. Both spin Laue and anti-spin Laue groups encode the parity rules enforced by the spin-only group. Using this framework, we establish three distinct classes of anti-spin Laue groups, $\mathbf{R}_\mathrm{s}^\mathrm{I}$, $\mathbf{R}_\mathrm{s}^\mathrm{II}$, and $\mathbf{R}_\mathrm{s}^\mathrm{III}$. Together with their even-parity counterparts \cite{Smejkal2021a}, they provide a comprehensive classification of collinear and coplanar magnetic systems with collinear momentum-space spin polarization, which is summarized in Tab. \ref{tab:spin_laue_group_types}.

The 21 anti-spin Laue groups identified in Table~\ref{tab:onecolumn_spin_splitting} combined with the 10 spin Laue groups of Ref.~\cite{Smejkal2021a} provide the complete momentum-space partial-wave classification of the unconventional magnets described by $\mathbf{R}_\mathrm{s}^\mathrm{III}$.
Since all coplanar systems with nonzero propagation vector ($\boldsymbol{q}\neq\boldsymbol{0}$) exhibit collinear momentum-space spin polarization, each of the 5748 spin space groups (SSGs) of Ref.~\cite{Xiao2024} are unambiguously assigned to one of these three classes. In particular, for  anti-altermagnets, this equates to reducing 160 inequivalent SPGs to only 21 anti-spin Laue groups.

Building on this comprehensive momentum space picture we develop a systematic symmetry-guided procedure that constructs canonical four-band Bloch Hamiltonians across the entire classification. This model-building goes beyond existing approaches that have largely relied on the canonical model of Ref.~\cite{Hellenes2023} or on effective two-band theories only valid near the $\Gamma$ point~\cite{Ezawa2025a, Hirschmann2026}.

\noindent{\it Construction of anti-spin Laue groups} --
We consider coplanar magnets in which spin-translation symmetries enforce a collinear momentum-space spin texture, while the spin-only group fixes its parity-constraint. This constraint yields an odd-parity out-of-plane component and an even-parity in-plane spin polarization (see End Matter). Therefore, each spin point-group element $[R_s\Vert g]$ -- $R_s$ being a spin-space rotation and $g$ a real space operation -- preserves or reverses this component, defining the sign character $\chi_\perp([R_s\Vert g])=\pm1$. Using this sign character we project the full spin group onto the relevant action on the out-of-plane component of the spin polarization.
The map $\chi_\perp$ descends to a one-dimensional irreducible representation $\gamma_\perp$ of the spatial parent point group $G$, defined by $\gamma_\perp(g)= \chi_\perp([R_s\Vert g])$ (see SM for details).

We then define the anti-spin Laue group as the graph of this sign representation,
\begin{equation}
    \mathcal{L}_{\mathrm{anti}}
    =
    \Gamma_{\gamma_\perp}
    =
    \{(\gamma_\perp(g),g)\mid g\in G\}
    \subset
    \{\pm1\}\times G .
    \label{eq:anti_spin_laue_definition}
\end{equation}
To connect to spin group notation, an element with $\gamma_\perp(g)=+1$ is written as $[E\Vert g]$, while an element with $\gamma_\perp(g)=-1$ is written as $[C_2\Vert g]$, in order to connect to spin group theory again. Here, $C_2$ denotes any spin-space two-fold rotation that reverses the normal spin component; after the reduction, only this binary action is retained. In the Hermann-Mauguin notation, we indicate the latter by a superscript ``$2$'' on the corresponding spatial operation. Thus, the superscript ``$2$'' denotes operations that reverse $S_{\perp}(\vec{k})$. 

Contrary to the construction of spin Laue case in which one adds an effective inversion symmetry \cite{Smejkal2021a}, for anti-spin Laue groups one must add an effective "anti-inversion" symmetry; in our notation given by $[C_2||\mathcal{P}]$.

\begin{figure}[t!]
    \centering
    \includegraphics[width=\linewidth]{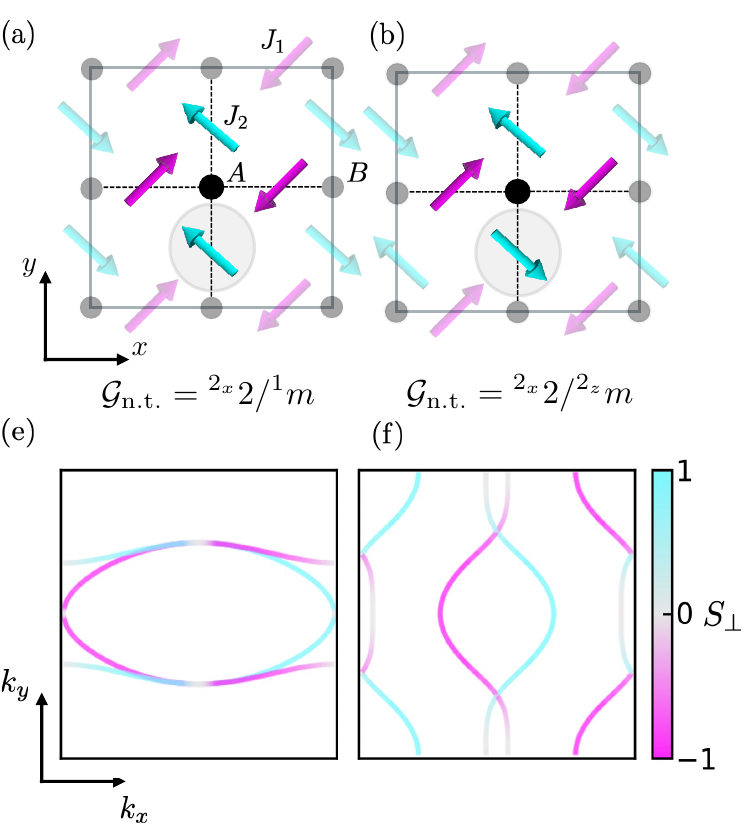}
    \caption{\textbf{Different real-space lattice models belonging to the same anti-spin Laue group ${}^22/{}^1m$.}
    (a),(b) Real-space lattice models with coplanar magnetic textures described by the nontrivial spin point groups $\mathcal{G}_{\mathrm{n.t.}} = {}^{2_x}2/{}^1m$ and $\mathcal{G}_{\mathrm{n.t.}} = {}^{2_x}2/{}^{2_z}m$, respectively. 
    (c),(d) Isoenergy lines of the corresponding models in momentum space. $S_\perp$ denotes the out-of-plane spin polarization of the respective bands. Despite their distinct real-space spin symmetries, both exhibit the same nodal structure, with a nodal line at $k_x=0$.
    }
    \label{fig:different_p_wave_models}
\end{figure}

Similar to the case of spin Laue groups~\cite{Smejkal2021a}, we distinguish three cases for anti-spin Laue groups:
$\mathbf{R}_\mathrm{s}^\mathrm{I} = [E||G], \mathbf{R}_\mathrm{s}^\mathrm{II} = [E||G]+[C_2||G]$, 
and  $\mathbf{R}_\mathrm{s}^\mathrm{III} = [E||H]+[C_2||G-H]$, where $H$ is a halving subgroup of $G$. {The case of  $S_\parallel$ being the only non vanishing component is described by spin Laue groups}, for it is constrained to be even-parity by the spin only group. In the case of anti-spin Laue groups, due to the addition of the effective anti-inversion  $[C_2||\mathcal{P}]$ there is no type I case. As in the even-parity case, $ \mathbf{R}_\mathrm{s}^\mathrm{II}$ describes system with spin degeneracy accross the entire Brillouin Zone, due to the presence of the $[C_2||E]$ element. The third class, $\mathbf{R}_\mathrm{s}^\mathrm{III}$, is defined by a noncentrosymmetric halving subgroup $H$ of the spatial parent, in contrast to spin Laue groups, which require $H$ to be centrosymmetric. This unconventional compensated phase exhibits nodal odd-parity spin splitting with  $p$-, $f$-, $h$-, $j$-, and $l$- partial-wave character.

There are 21 distinct anti-spin Laue groups shown in Tab. \ref{tab:onecolumn_spin_splitting}, which compose the counterpart to the 10 spin Laue groups describing the even-parity $\mathbf{R}_\mathrm{s}^\mathrm{III}$.

We also show material candidates for anti-altermagnets for the respective anti-spin Laue groups if available and refer to the SM for a comprehensive MAGNDATA screening of anti-altermagnetic materials \cite{MAGNDATA,Gallego2016}.

\noindent{\it Many-to-one reduction to anti-spin Laue symmetry} --
We now demonstrate a key advantage of the anti-spin, and spin Laue group concept: it makes explicit the many-to-one reduction from distinct coplanar spin point groups (SPGs) to a common momentum-space symmetry class. Projecting all coplanar SSGs onto inequivalent SPGs yields 252 distinct groups~\cite{Schiff2025a}, of which 160 are compatible with odd-parity magnetism/anti-altermagnetism in the presence of the corresponding spin-translation symmetry and are counted in Tab.~\ref{tab:onecolumn_spin_splitting}. \footnote{Note that the coplanar count does not correspond to the purely algebraic Litvin SPGs \cite{Litvin1974,Litvin1977}, but rather to all nonequivalent combinations of a given Litvin SPG with the coplanar spin only group \cite{Schiff2025a}(see SM for more details).}. An analogous reduction underlies spin Laue groups in which the spin polarization is always even-parity. A key difference between the collinear and coplanar case when it comes to this reduction is the fact that the only allowed nontrivial spin translation symmetry in the collinear case is $[C_{2\perp}||\{E|\boldsymbol{t}\}]$, where coplanar systems can also exhibit generic spin helix structures with a $n$-fold rotation around the axis perpendicular to the magnetic plane. This additional complexity leads to 5748 commensurate coplanar SSGs that can also be classified within the three distinct classes of either spin Laue or anti-spin Laue groups, depending on which component of the spin polarization remains.

We now illustrate the conceptual value of this reduction through a model example and a concrete material. Figure~\ref{fig:different_p_wave_models} shows schematics of two lattice systems with two nonequivalent coplanar SPGs that realize the same $p$-wave anti-spin Laue class. Restricting the spin-space operations of each model to their action on the perpendicular spin direction $S_\perp(\boldsymbol{k})$, determines the anti-spin Laue group to be ${}^22/{}^1m$. However in real space, the non-trivial SPGs differ substantially.

\begin{figure}[t!]
    \centering
    \includegraphics[width=\linewidth]{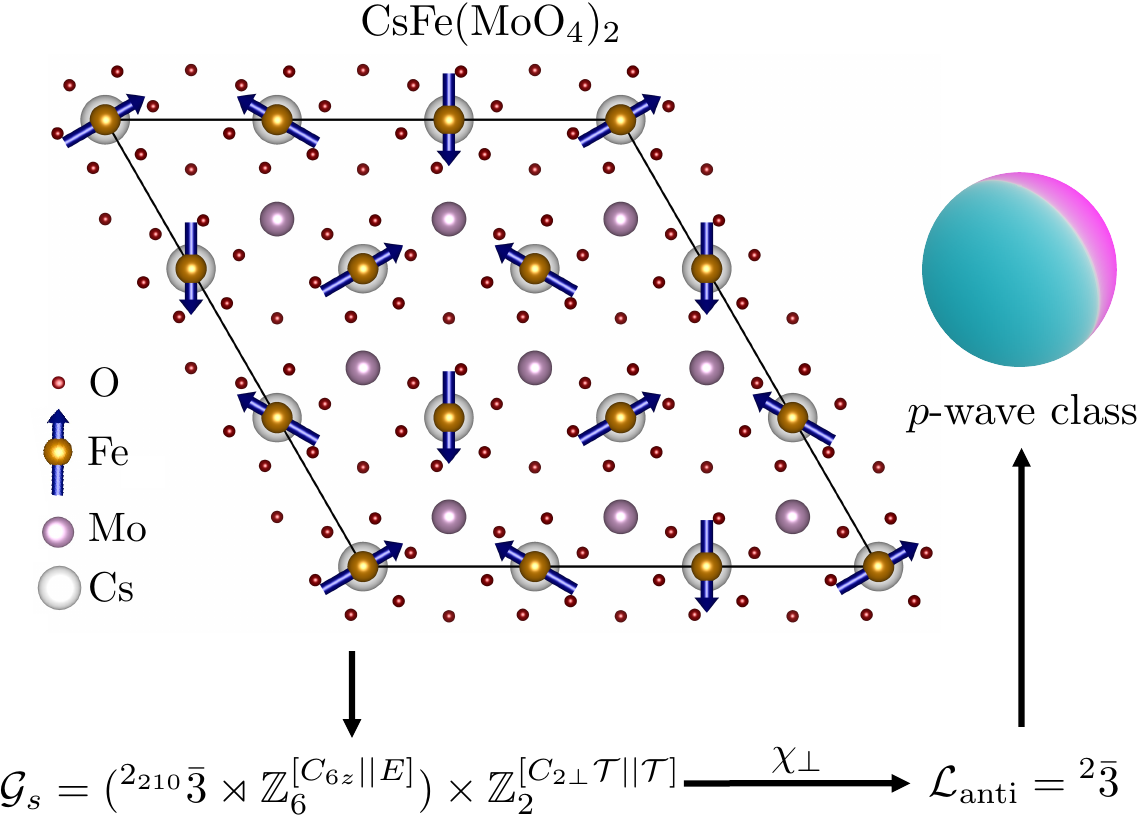}
    \caption{\textbf{Reduction of complex SSG to concise anti-spin Laue group in CsFe(MoO$_4$)$_2$.}
    A top view of the spin helix structure of CsFe(MoO$_4$)$_2$ and its spin point group is indicated. $\chi_\perp$ denotes the anti-spin Laue group reduction to the the group ${}^2\bar{3}$ which is in the $p$-wave class of Tab. \ref{tab:onecolumn_spin_splitting}.}
    \label{fig:different_p_wave_models_material}
\end{figure}

The canonical minimal model introduced for CeNiAsO in Ref.~\cite{Hellenes2023} and used repeatedly throughout the literature~\cite{Brekke2024,Chakraborty2025a,Hellenes2023,Mitscherling2026} has the nontrivial SPG

given by the Litvin symbol ${}^{2_x}2/{}^1m$. The corresponding real-space magnetic structure is shown in Fig.~\ref{fig:different_p_wave_models}(a). Figure~\ref{fig:different_p_wave_models}(b) shows a lattice model with a different nontrivial SPG, namely ${}^{2_x}2/{}^{2_z}m$. Notably, the spin parent group (see End Matter) of this group is dihedral (222), i.e. has 3 inequivalent perpendicular spin rotation axis. This difference between the two models is highlighted in the magnetic environment of the A sublattices in Fig.~\ref{fig:different_p_wave_models}(a) and Figure~\ref{fig:different_p_wave_models}(b). Nevertheless, restricting both spin actions to the perpendicular spin direction $S_\perp(\boldsymbol{k})$ gives the same sign representation of the spatial parent and hence the same anti-spin Laue group ${}^22/{}^1m$.

The common anti-spin Laue class fixes the leading momentum dependence of the spin polarization, $S_\perp(\boldsymbol{k})\sim k_x$, and therefore enforces a spin-nodal line at $k_x=0$. This common momentum-space symmetry is visible in the iso-energy lines in Figs.~\ref{fig:different_p_wave_models}(c) and \ref{fig:different_p_wave_models}(d). This example demonstrates how distinct SPGs reduce to the same anti-spin Laue group, thus sharing the same symmetry and nodal structure of $S_\perp(\vec{k})$.
\begin{figure*}[t!]
    \centering
    \includegraphics[width=1\linewidth]{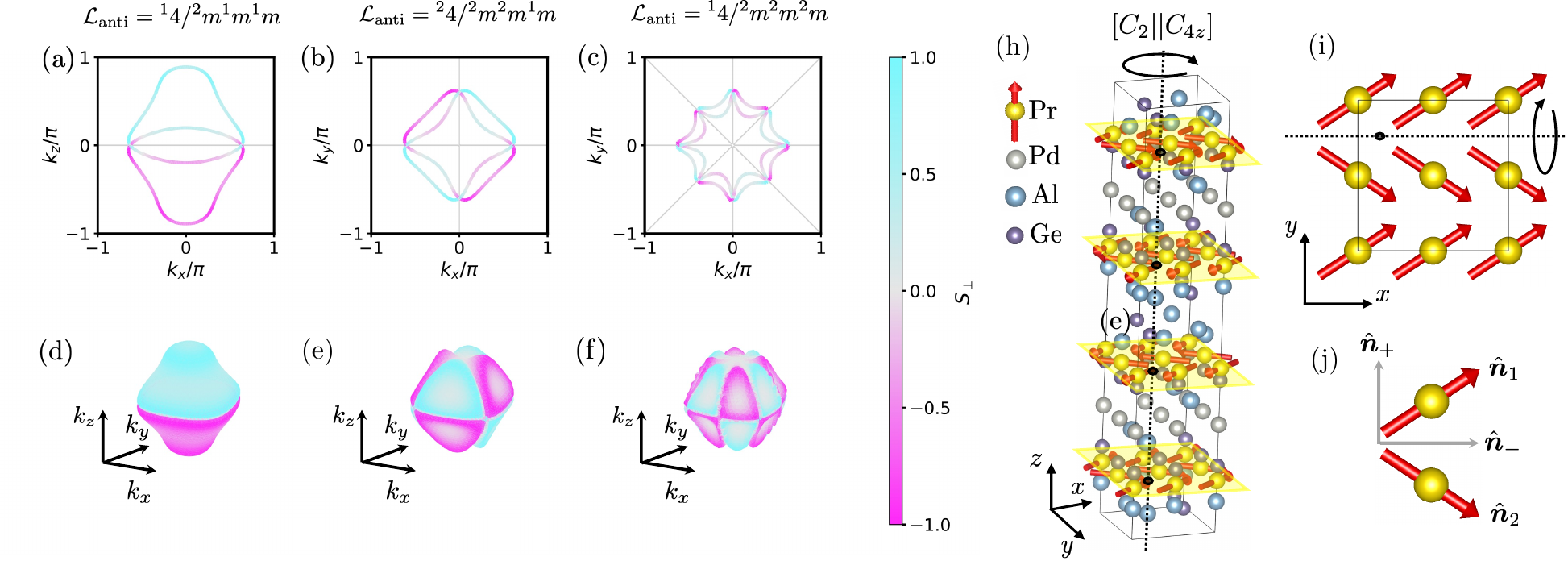}
    \caption{\textbf{Type-A models for three anti-spin Laue groups of the $4/mmm$ crystal system.}
    (a)--(c) Spin-polarized isoenergy cuts of the $p$-, $f$-, and $h$-wave representatives with anti-spin Laue groups ${}^14/{}^2m{}^1m{}^1m$, ${}^24/{}^2m{}^2m{}^1m$, and ${}^14/{}^2m{}^2m{}^2m$, respectively. (d)--(f) Corresponding three-dimensional isoenergy surfaces. The color code denotes $S_\perp$ and reveals the symmetry-enforced nodal planes. (h)--(j) $f$-wave candidate Pr$_2$PdAl$_7$Ge$_4$~\cite{Gao2023b}: its physical moments are interchanged by the $[C_2||C_4]$ spin-symmetry, whereas their symmetric and antisymmetric combinations form a symmetry-adapted basis $\boldsymbol{\hat{n}_\pm}$.}
    \label{fig:minimal_models_electronic_bandstructure}
\end{figure*}

To demonstrate the directness of the anti-spin Laue-group description for real materials, we consider the magnetic crystal structure of CsFe(MoO$_4$)$_2$ \cite{Gagor2014}, shown in Fig.~\ref{fig:different_p_wave_models_material}. The material possesses a rather complex SPG, arising from the spin-translation group associated with its sixfold spin helix and a nontrivial trigonal SPG. Consequently, the nodal structure of the out-of-plane spin polarization is not immediately readable from the group structure. After reduction, however, the anti-spin Laue group takes the remarkably concise form ${}^{2}\bar{3}$, which Table~\ref{tab:onecolumn_spin_splitting} identifies as $p$-wave. This example illustrates how intricate real-space spin textures can reduce to simple momentum-space symmetry classifications, highlighting the utility of the anti-spin Laue-group framework, that also applies to the spin helix ferroelectric $p$-wave magnet NiI$_2$ \cite{Song2025a}.

Likewise, previously proposed low-energy models are naturally organized by the same anti-spin Laue classes, placing distinct material realizations and their effective descriptions within a common symmetry framework~\cite{Hirschmann2026,Ezawa2025a}.

\noindent{\it Minimal models from anti-spin Laue symmetry} --
Besides providing a comprehensive taxonomy of momentum-space spin textures, anti-spin Laue groups yield a direct recipe for constructing canonical lattice four-band representatives, analogous to the spin-Laue construction for altermagnets in Ref.~\cite{Roig2024}.

For this, we focus on minimal spinful two-sublattice models for coplanar anti-altermagnets. We adopt a canonical basis in which the two magnetic sublattices are exchanged only by the spin translation, $\mathcal{T}\vec{t}=\tau_x\otimes i\sigma_y K$, where $\tau_i$ and $\sigma_i$ act in sublattice and spin space, respectively. Even when the paramagnetic Wyckoff multiplicity is higher than one, and this assumption does not hold, we retain this minimal constraint for the exchange of the sublattices, since the resulting four-band model is intended as a low-energy symmetry representative and can still reproduce the qualitative band structure and spin splitting of the full material calculation, as demonstrated for CeNiAsO \cite{Mitscherling2026, Chakraborty2025a}. 

The more relevant information is how the spin space operation transforms the set of moments in a particular material. This information is given by the spin parent and its action on the magnetic moment directions.

We choose to describe the exchange interaction mediated by the magnetic moments on the system by spin dependent hoppings, and restrict to models with two separate physical moment directions $\hat{\boldsymbol{n}}_1$ and $\hat{\boldsymbol{n}}_2$. Following Ref.~\cite{Mitscherling2026} the out-of-plane spin polarization for such a 4 band model is then given by
\begin{equation}
    S_{\perp}(\vec{k})
    \sim
    d_1(\vec{k})
    \left[
        \vec{d}_2(\vec{k})
        \times
        \vec{d}_3(\vec{k})
    \right]_{\perp},
    \label{eq:spin_cross_product_prl}
\end{equation}
where $\vec{d}_2(\vec{k})$ and $\vec{d}_3(\vec{k})$ are the two spin dependent hoppings connected to $\tau_2\boldsymbol{\sigma}$ and $\tau_3\boldsymbol{\sigma}$ respectively. The magnetic moment directions and their associate spin-dependent hopping mediating the exchange interaction are encoded in $\vec{d}_2(\vec{k})$ and $\vec{d}_3(\vec{k})$. The $\mathcal{T}\boldsymbol{t}$ symmetry enforces $\vec{d}_2(\vec{k})$ to be composed of odd and $\vec{d}_3(\vec{k})$ of even momentum functions (see SM).

The anti-spin Laue group directly specifies the one-dimensional sign irrep $\Gamma_{S_\perp}$ carried by the perpendicular spin polarization.
Equation~\eqref{eq:spin_cross_product_prl} thus converts the construction of anti-altermagnetic models into a representation-theoretic problem. The three Hamiltonian sectors must therefore satisfy the irrep-product rule $\Gamma_{d_1}
    \otimes
    \Gamma_{\boldsymbol{d}_2}
    \otimes
    \Gamma_{\boldsymbol{d}_3}
    =
    \Gamma_{S_\perp}.$
    
In the following, we restrict $d_1$ to the trivial irrep of the point group and distinguish three minimal four-band construction Types.

In Type-A models, the full anisotropy of the spin polarization is carried by the odd sector $\vec{d}_2$ of the Hamiltonian, while $\vec{d}_3$ transforms trivially. In Type-B models, both $\vec{d}_2$ and $\vec{d}_3$ transform nontrivially and combine to yield $\Gamma_{S_\perp}$. A third possibility arises when the magnetic moments, and hence the associated Pauli spin matrices, transform according to a genuine two-dimensional irrep of the spin parent group. We neglect this Type-C case in this Letter and note that it occurs for systems with spin parent groups $622$, $422$, and $C_n$ with $n>2$. Together with the assumption that only $\mathcal{T}\boldsymbol{t}$ exchanges the sublattices, this information fixes the irreps of $\vec{d}_2$ and $\vec{d}_3$.

To construct these one-dimensional irreps in Type-A and Type-B models, the moments encoding the exchange interaction in the Hamiltonian must be expressed in a symmetry-adapted basis of the spin parent rather than the explicit physical moment directions $\hat{\boldsymbol{n}}_{1,2}$ (see SM). We therefore write $\vec{d}_2=\hat{\boldsymbol{e}}_- d_2'(\boldsymbol{k})$ and $\vec{d}_3=\hat{\boldsymbol{n}}_+ d_3'(\boldsymbol{k})$, where $\hat{\boldsymbol{n}}_\pm$ are symmetry-adapted basis vectors carrying well-defined sign characters, such that the spin-space operations act as $\pm1$ on the corresponding moment directions. For example, in a Type-A model with two different physical moment directions, either $\hat{\boldsymbol{n}}_{1,2}$ already transform according to two distinct one-dimensional irreps of the spin parent, or one constructs the symmetry-adapted combinations $\hat{\boldsymbol{n}}\pm \sim \hat{\boldsymbol{n}}_1 \pm \hat{\boldsymbol{n}}_2$.

This sign character identifies which spin group elements reverse a given symmetry-adapted moment direction and, consequently, which spatial operations are paired with them. The corresponding momentum functions $d_2'$ and $d_3'$ must then acquire the compensating sign under these spatial operations, thereby fixing their point-group irreps (see SM for details).
We now focus on the Type-A construction case in the crystal structure $4/mmm$, which yields a representatives for $p$,$f$ and $h$-wave anti-altermagnets. In particular, the $f$-wave model we construct respects the spin-symmetries of the candidate material Pr$_2$PdAl$_7$Ge$_4$ \cite{Gao2023b}, which has two different axes along which the magnetic moments point. The corresponding anti-spin Laue groups are given by ${}^14/{}^2m{}^1m{}^1m$, ${}^24/{}^2m{}^2m{}^1m $ and ${}^14/{}^2m{}^2m{}^2m$, for $p$-, $f$- and $h$-wave anti-altermagnetism respectively. In Fig. \ref{fig:minimal_models_electronic_bandstructure}(a)-(c) we show isoenergy surface cuts for the three models. In Fig.\ref{fig:minimal_models_electronic_bandstructure}(d)-(f) the the full isoenergy surface reveals all nodal planes of the spin polarization (see SM for details of the model Hamiltonians). Figure \ref{fig:minimal_models_electronic_bandstructure}(h) shows a unit cell of  Pr$_2$PdAl$_7$Ge$_4$ with the $C_4$-rotation axis indicated, which is accompanied by a $180^\circ$ rotation in spin space. Figues \ref{fig:minimal_models_electronic_bandstructure}(i) and (j) illustrate how the physical moment directions are interchanged under the spin-space rotation, while the symmetry-adapted combinations $\hat{\boldsymbol{n}}_\pm$ transform with well-defined sign characters. The phenomenological model can thus be constructed in the symmetry-adapted basis and subsequently mapped back onto the physical moment directions.

\noindent{\it Conclusions}-- We introduced anti-spin Laue groups as the reduced momentum-space symmetry description of coplanar magnets with odd-parity collinear spin polarization. In analogy to the classification of collinear magnetic phases, which lead to the discovery of altermagnetism \cite{Smejkal2021a}, the unconventional third class of anti-spin Laue groups $\mathbf{R}_s^{\mathrm{III}}$ classifies all possible antialtermagnets into well delimited 21 anti-spin Laue groups. The sign representation used to build the anti-spin Laue groups generates a many-to-one reduction of 160 SPG to only 21 ASLGs. Furthermore, our analysis demonstrates that every coplanar magnet with nonzero propagation vector will have collinear spin polarization in momentum space. Our formalism, is therefore suitable to a broad family of materials. By screening MAGNDATA, we identified the anti spin Laue group of 83 realistic antialtermagnetic candidates. While our main focus is on coplanar magnets with collinear spin polarization, by assigning one anti spin Laue group to each spin component, our framework can be easily extended to systems with mixed parity among multiple spin components. Our classification places odd- and even-parity unconventional magnets on equal footing, and provides a direct symmetry-guided route for building canonical four-band models for antialtermagnetism.

\paragraph{Acknowledgements}This work was funded by the German Research Foundation (DFG) through TRR 173-268565370 (Projects No.~A03 and B13), TRR 288-422213477 (Projects No.~A09 and B05), and Project No.~504261060 (Emmy Noether Programme). We acknowledge support by the Dynamics and Topology Center (TopDyn) funded by the State of Rhineland-Palatinate. We acknowledge the high-performance computational facility of supercomputer “Mogon” at Johannes Gutenberg-Universität Mainz, Germany.


%

\appendix

\section*{END MATTER}

\section{Spin group notation and parity constraints}

In the main text we have considered coplanar magnets with a collinear spin polarization in momentum space. This spin polarization is produced by the presence of a spin-translation symmetry $[R_s||E|\boldsymbol{t}]\, ,\, R_s\in \mathrm{SO}(3)_\mathrm{spin}$ \cite{Watanabe2024}. In momentum space, it acts on the spin expectation value as with point symmetry part $[R_s||E]$, which leads to vanishing components of spin polarization perpendicular to the spin-rotation axis of $R_s$. The group composed of such spin translations in which the zero translation $\mathbf{t}=0$ is only linked with the identity in spin space as $[E||E|0]$ forms the nontrivial spin translation group, that we denote as $\mathcal{G}_\mathrm{s.t.}$. In the main text, for readability, we sometimes omit nontrivial, and simply call it spin translation group. Throughout, symmetry operations on the left side of $\parallel$ act on spin space and elements on the right hand side on real space. 

The spin-only group $\mathcal{G}_\mathrm{s.o.}$ is the set of group elements of the spin space group, which are given by some operation in spin space and the identity in real space. The final part of the total spin group is the non-trival spin space (group) $\mathcal{G}_\mathrm{n.t.}$ which is built via the Litvin theorem \cite{Litvin1974,Litvin1977,Schiff2025a}. This part of the group is built from a so called spatial parent group and spin parent group \cite{Schiff2025a}. The spin parent group is the group of all spin space rotations present in the non-trivial spin space (point group). In total the entire spin group can be written as \cite{Watanabe2024}:
\begin{align}
    \mathcal{G}_s = (\mathcal{G}_\mathrm{n.t.}\rtimes\mathcal{G}_\mathrm{s.t.})\times \mathcal{G}_\mathrm{s.o.}
\end{align}

The spin-only group for coplanar systems is $\mathbb{Z}_2^{[C_{2\perp}\mathcal{T}||\mathcal{T}]}$, where $C_{2\perp}$ denotes a $180^\circ$ spin-space rotation about the axis normal to the coplanar spin plane, and $\mathcal{T}$ the time reversal operation. This time reversal operation acting in real space induces an effective inversion symmetry in momentum space. The simulantenous $C_{2\perp}\mathcal{T}$ in spin space leaves the in-plane spin polarization invariant and reverses the out-of-plane components, thus inducing the even- and opp-parity of $S_\parallel(\vec{k})$ and $S_\perp(\boldsymbol{k})$ constraints, respectively. Hence the symmetry of the even-parity in-plane spin polarization is described by adding an effective inversion to the SPG, yielding the well known spin-Laue groups, which are also used to describe altermagnets \cite{Smejkal2021a}.
This generalization of the application of spin Laue groups opens the possibility of noncollinear systems with collinear $d$-,$g$- and $i$-wave spin polarization in momentum space \cite{Song2025d}.

The symmetries of the odd-parity out-of-plane spin polarization are therefore obtained by extending the projected spin-point-group action on $S_\perp$ with the effective anti-inversion $[C_2\Vert\mathcal{P}]$, yielding the anti-spin Laue groups. In the presence of  a $\mathcal{T}\boldsymbol{t}$ symmetry, the anti-spin Laue groups therefore comprehensively classify  anti-altermagnetism \cite{Hellenes2023,Jungwirth2025a}.
While our main focus in this Letter is on coplanar magnets with collinear spin polarization, by assigning one anti spin Laue group to each surviving spin component, our framework can be easily extended to systems with mixed parity among multiple spin components \cite{Pari2025, Ryu2026}.

\end{document}